\documentclass[journal, twocolumn]{IEEEtran}
\IEEEoverridecommandlockouts
\usepackage[ruled,linesnumbered,vlined]{algorithm2e}
\usepackage{multirow}
\usepackage{makecell}
\usepackage{caption}
\usepackage{graphicx}
\usepackage{float}
\usepackage{subfigure}
\usepackage{pdfpages}
\usepackage{tabularx}
\usepackage{amsmath}
\usepackage{amsfonts,amssymb}
\usepackage{stmaryrd}
\usepackage{diagbox}
\usepackage[ruled]{algorithm2e}
\newtheorem{problem}{\textbf{Problem}}
\newtheorem{definition}{\textbf{Definition}}
\newtheorem{example}{\textbf{Example}}
\newtheorem{proposition}{\textbf{Proposition}}

\usepackage{cite}
\usepackage{bm}
\usepackage{xcolor}

\usepackage{CJK}
\usepackage{indentfirst}
\usepackage{cases}
\usepackage{subeqnarray}
\usepackage{lettrine}

\begin{document}

\title{Scheduling of Flexible Manufacturing Systems Based on Place-Timed Petri Nets and Basis Reachability Graphs}

\author{Zhou He,~\IEEEmembership{Senior Member,~IEEE}, Ning Li, Ning Ran,~\IEEEmembership{Member,~IEEE}, and Liang Li \thanks{This work was supported in part by the National Natural Science Foundation of China under Grant nos. 62373234, 62373132, and 62303359; in part by the Shaanxi Provincial Natural Science Foundation under Grants 2023-JC-YB-564 and 2022JM-330; in part by the Foundation of Hebei Education Department under Grant BJ2021008; and in part by the Chunhui Project of Ministry of Education of China under Grant HZKY20220257. (Corresponding author: \emph{Liang Li})} \thanks{Zhou He and Ning Li are with the School of Electrical and Control Engineering, Shaanxi University of Science and Technology,
Xi'an 710021, China. (email: hezhou@sust.edu.cn; 230612026@sust.edu.cn)} 
\thanks{Ning Ran is with the College of Electronic and Information Engineering, Heibei University, Baoding 071002, China. (email: ranning87@hotmail.com)}%
\thanks{Liang Li is with the School of Artificial Intelligence and Automation, Wuhan University of Science and Technology, Wuhan 430081, China (e-mail: liangli@wust.edu.cn)}
}

\maketitle
\thispagestyle{empty}
\pagestyle{empty}

\begin{abstract} 
Scheduling is a key decision-making process to improve the performance of flexible manufacturing systems. Place-timed Petri nets provide a formal method for graphically modeling and analyzing such systems. By generating reachability graphs and combining intelligent search algorithms, operation sequences  from the initial state to the target state can be found for the underlying system. However, the reachability graph grows exponentially with the system size increases, which is the main challenge of existing methods for scheduling large systems. To this end, we develop an efficient improved  beam search algorithm  to optimize the makespan based on a compact representation of reachability graph called basis reachability graph. The key idea behind the proposed method is to form a state together with the basis markings and its corresponding transition sequences, and  evaluate the cost of the state based on the resource idle time. Experimental results are conducted on several benchmark systems which show that the developed method improves the search efficiency while ensuring the quality of the solution compared with existing methods.


\end{abstract}


\IEEEpeerreviewmaketitle
\section{Introduction} \label{1}
\IEEEPARstart{F} {lexible} manufacturing systems (FMSs) have become a common production model in modern manufacturing, widely used in industries such as  aerospace and consumer electronics. An FMS significantly improves the adaptability of the manufacturing system by providing flexible production capacity and diversified machining paths. In FMSs, scheduling needs to arrange the operation time and assign the resource of each process step under some production constraints to optimize some criteria, e.g., makespan, cost, energy, and so on.
The constraints, including operation precedence, shared resources, machine capability, route diversity, make the scheduling problem complicated \cite{Li2014}.

The scheduling of FMSs belongs to one of the NP-hard combinatorial problems \cite{Fanti2016}. To address this problem, researchers have proposed a variety of methods, including mathematical planning \cite{Pan2018,Kammoun2017,Qiao2019,Yang2020}, neighborhood search \cite{Abdelmaguid2015}, heuristic scheduling methods \cite{KeYiXing2012,Cao2019}, and artificial intelligence techniques \cite{Luo2024,Hu2020}.
Although each of these methods has certain advantages and  application scopes, there are some limitations in practical applications.
Mathematical planning is suitable for small scale problems, while in large-scale scenarios, it is limited by high complexity and real time demand.
Heuristic algorithms are robust, but easy to fall into the local optimum.
Neighborhood search methods can jump out of the local optimum, while the convergence speed is slow and relies on the initial solution.


An FMS is a standard discrete event system (DES) whose components and processes can be viewed as discrete states and events (such as equipment start-up, task completion, and material handling operations) \cite{ZhiWuLi2012}. Petri nets (PNs), as a modeling, analysis, and monitoring tool for DES \cite{Mazi2022,He2021,Li2019,Cherif2021}, have been extensively applied in the deadlock control, performance analysis, and scheduling of FMSs. 
For instance, the authors in \cite{Li2019} develop a modified dynamic programming model for the  scheduling problem of PNs to avoid the inability to obtain optimal or suboptimal scheduling in an acceptable time due to the exponential growth of the number of states with the problem size.
The scheduling problem of FMSs in uncertain environments is investigated in \cite{Cherif2021}. An improved beam search algorithm is proposed to selectively explore the state space of the PN model, which ensures the quality of scheduling while improving the solution efficiency.


In this paper, we propose a heuristic scheduling method of FMSs to optimize the makespan using place-timed PNs. 
We first construct a place-timed PN system for scheduling of an FMS which responds to the system behavior and manufacturing specifications.
We then develop a heuristic algorithm to search the basis reachability graph (BRG) space of this system, which is a compact representation of the reachability graph (RG) of the PNs that preserves the basic structural and behavioral information of the original system. In the heuristic search phase,  a generation filtered beam search algorithm is developed based on the resource idle time that evaluates and compares states based on the underlying identifiers and their corresponding transition sequences.
Extensive experimental studies on benchmark  systems demonstrate the superiority of the proposed method in terms of solution quality and computational efficiency compared to the existing approaches.


This paper is divided into following sections. Section \ref{state} introduces the state of the art. Section \ref{Preliminary} describes the necessary preparatory work, which includes some of the theoretical foundations required for the study of this paper. The scheduling algorithm based on the BRG is elucidated in Section \ref{4}. The comparative results and simulation experiments on benchmark systems are discussed in Section \ref{5}. Finally, conclusion and future work are presented in Section \ref{conclusion}.

\section{State of the art}\label{state}

Early scheduling methods based on untimed PNs can effectively model the structural properties of FMSs but failed to incorporate critical temporal constraints such as operation durations and resource delays. This limitation results in schedules that overlook practical issues such as resource contention and adherence to deadlines. To address these limitations, temporal dynamics are modeled using timed Petri nets, specifically transition-timed Petri nets and place-timed Petri nets.
Lee \emph{et al}. \cite{DooYongLee1994} are the first to apply the classical A$^{\ast}$ algorithm of graph theory, which incorporates a heuristic function to limit the search space. Later, several admissible heuristic functions are proposed for the A$^{\ast}$ search within the reachability graph of PNs to achieve optimal system schedules.
These works aim to direct the search toward the goal node by assigning greater weight to nodes deeper in the search tree, thereby avoiding state space explosion. Generally, most approaches extend the A$^{\ast}$ algorithm by limiting the nodes to explore or introducing new heuristic functions.
Xiong \emph{et al}. \cite{HuanxinHenryXiong1998} propose an admissible heuristic function that estimates the cost from the current state to the target state by calculating the maximum sum of operation times for the remaining operations on each machine. The work in \cite{Huang2012} presents a near admissible heuristic search strategy for FMS scheduling with alternative routings. Building on this, the authors develop an admissible heuristic function based on place-timed Petri nets and an enhanced dynamic weighting A$^{\ast}$ algorithm.
Huang \emph{et al}. \cite{Huang2014} proposes an improved search strategy that combines admissible and non-admissible heuristic functions within the A$^{\ast}$ algorithm in reachable graphs for FMS scheduling.


The scheduling of FMSs based on timed Petri nets (TPNs) faces critical challenges such as state space explosion and computational inefficiency in large-scale systems. Some works use selective strategies to reduce the number of nodes explored to improve the efficiency. 
{The idea of model predictive control  is introduced to limit the search range. The work in \cite{Lefebvre2016} dynamically adjust the prediction range and sequence length according to the distance between the current marker and the target marker, and search for the control sequence by partially exploring the RG.
An efficient deadlock-free scheduling algorithm based on dynamic windowing strategy is proposed by \cite{JianChaoLuo2015}, which significantly reduces the search space and improves the scheduling efficiency by restricting the local search space.}
The branch and bound algorithm for enumerating some of the candidate solution nodes and pruning them by tight lower bounds and dominance rules to improve the algorithmic performance is presented in \cite{jeong2024}.
The integration of the beam search with TPN is investigated in recent studies \cite{Meja2009, Meja2005, Meja2016, Lefebvre2018, Lefebvre2021}. Lefebvre \emph{et al}. \cite{Lefebvre2018} achieve robust scheduling by designing a cost function that incorporates risk assessment and a beam search algorithm to ensure that the system reaches the target state with minimal time and risk. In \cite{Lefebvre2021} a coding method of TPN is proposed to accurately estimate the remaining process time in the production plant using a tree structure and linear matrix inequalities. Its effectiveness in scheduling optimization is verified in combination with the beam search algorithm.

Other methods combined meta-heuristic algorithms and TPNs to schedule FMSs. The work in \cite{Meja2012} introduces a genetic algorithm employing a chromosome structure with a transition conflict-resolution mechanism alongside a rejection strategy to eliminate infeasible solutions.
Xing \emph{et al}. \cite{KeYiXing2012} develop a search approach based on genetic algorithm, which integrates various standard deadlock avoidance supervisors.
{In \cite{Nouiri2015}, Nouiri \emph{et al}. propose a multi-agent optimization algorithm based on particle swarm optimization through multiple intelligences collaborating, communicating, and self-adjusting in the solution space to solve a flexible job-shop problem (FJSP).
Recently, population-based meta-heuristic algorithms become important tools in FJSP optimization problems.
Xing \emph{et al}. introduce knowledge into the ant colony optimization algorithm to solve the problem. The use of discrete gray wolf algorithm in \cite{Jiang2018} effectively solves the scheduling problem in FJSP. In addition, there exist many similar optimization algorithms,} including the Jaya algorithm \cite{Caldeira2019}, brain storm optimization \cite{Alzaqebah2022}, and the dragonfly algorithm \cite{Yang2022}.

Our approach is different from existing methods that explore the whole RGs of TPNs. Instead, we use a compact representation of RGs that preserves the basic structural and behavioral information of the original system to search for the optimal schedules.  Meanwhile, the computational complexity of A$^{\ast}$ algorithms typically grows significantly with system size, making it challenging to obtain a solution within a reasonable time frame. In this paper, we develop an improved beam search algorithm to explore optimal solutions based on the BRGs of TPNs, which is simpler and more efficient for FMSs containing a large number of workpieces to be machined and complex machining paths. The proposed method offers key advantages, including high accuracy and low computational effort, as the BRGs retain the structural and behavioral information of the original system while significantly reducing its state space.

\section{System scheduling with timed PNs}\label{Preliminary}

\subsection{Petri net}
A PN is composed of four-tuple $N = (P, T, Pre, Post)$, where $P$ is a finite set of $m$ \emph{places}; $T$ is a finite set of $n$ \emph{transitions}; $Pre: P \times T \rightarrow \mathbb{N}$ and $Post: P \times T \rightarrow \mathbb{N}$ are respectively \emph{pre-} and \emph{post-incidence function} that specify the arcs in the net, where $\mathbb{N} = \{0, 1, 2, \ldots\}$.
Specially, places and transitions are graphically represented by circles and bars, respectively.
The incidence matrix is defined as $C = Post - Pre \in \mathbb{Z}^{n\times m}$, where $\mathbb{Z} = \{0, \pm1, \pm2, \ldots\}$ denotes the set of integers.

A \emph{marking} is a function $M : P \rightarrow \mathbb{N}$ that is represented by an $m$-component vector $M \in \mathbb{N}^{m}$, where $\mathbb{N}$ denotes the set of non-negative integers. The count of tokens in place $p$ within marking $M$ is represented by $M(p)$.

A transition $t$ is said to be \emph{enabled} at a marking $M$ if $M \geq Pre(\cdot, t_i)$, where $Pre(\cdot, t_i)$ is the $pre$-incidence vector of $t_i$. Specifically, $Pre(\cdot, t_i)$ represents the weights of the input arcs from each place $p$ to the transition $t_i$. For a place $p$, if there exists an arc directed from $p$ to $t_i$ with weight $x$, then $Pre(p, t_i) = x$; otherwise, $Pre(p, t_i) = 0$. The condition $M \geq Pre(\cdot, t_i)$ ensures that there are enough tokens in each input place of $t_i$ to satisfy its firing requirements. This situation is denoted as $M[t_i\rangle$. If $t_i$ is enabled at $M$, then
it may fire and reach a new marking
\begin{equation}
\label{eq0}
M' = M + C(\cdot, t_i)
\end{equation}
denoted by $M[t_i \rangle M'$, where $C(\cdot, t_i)$ is the column vector of the incidence matrix $C$ corresponding to $t_i$. The set of all finite transition sequences over $T$ is denoted by $T^{\ast}$. Then $M[\sigma \rangle M'$ denotes that a transition sequence $\sigma = t_1t_2\ldots t_n \in T^{\ast}$ is enabled at $M$ and it's occurrence yields $M'$.
The \emph{firing vector} of $\sigma \in T^{\ast}$ is denoted as $\mathbf{y}_\sigma$. It holds $\mathbf{y}_{\sigma}(t) = y$ if transition $t$ occurs $y$ times in $\sigma$.
The set of all markings that can be reached from $M_0$ is denoted by $R(N, M_0)$. A pair $(N, M_0)$ is referred to as a \textit{PN system}.


The \emph{input} places and \emph{output} places of a transition $t \in T$ are indicated by $^{\bullet}{t}$ and ${t}^{\bullet}$, respectively. An analogous definition is given to the input transitions $^{\bullet}{p}$ and the output transitions ${p}^{\bullet}$ of place $p \in P$.


\subsection{Timed PN system for FMS scheduling}



In an FMS, each job can start at time 0 and no preemption is allowed.
The processing time for each workpiece on the corresponding resource is pre-set.
All processes of a workpiece must be executed in strict sequence, and any out of sequence operation will result in the scheduling program failing, thus affecting the feasibility and efficiency of the overall production plan.
Once an operation starts, it continues without interruption until it finishes.
Upon completing an operation, the job releases its current resource and proceeds to the next required one, provided that resource is available.
If not, the job remains at the current resource until the next one becomes free.
The scheduling problem in an FMS involves not only determining the sequence of operations for all jobs but also assigning appropriate routes for each job during processing.
The objective is to find a feasible schedule to minimize the latest of all workpieces to finish machining, i.e., \emph{makespan}.

For each processing route, a PN sub-model is constructed to accurately describe its dynamic behaviors such as state transfer, event triggering, and resource allocation. These sub-models can independently depict the processes of workpiece processing, resource utilization or material transportation. Finally, the sub-models are merged through a shared place to generate a complete system model.

The FMS considered in this paper comprises $q$ types of resources, which is denoted as $R = \{r_1, \dots, r_q\}$.
{A resource may represent a machining center, an automated guided vehicle, or a conveyor belt. The capacity of a resource $r_i \in R$ is denoted by $U(r_i)$, defined as the maximum number of jobs it can process simultaneously. A resource $r_i$ is considered available when it is either idle or processing no more than $U(r_i)$ jobs. The total capacity of all resources is given by $N_b^r = \sum_{i = 1}^q U(r_i)$.}

The system is capable of producing $s$ types of jobs, which is denoted as $B = \{b_1, \dots, b_s\}$.
Each job type has a specific lot size, where the lot size for job type $b_i \in B$ is denoted as $\varphi_i$, and the total number of jobs to be produced is $N_b^j = \sum_{i = 1}^s \varphi_i$.
Every job follows a predefined sequence of operations, referred to as its processing route.
A job may have multiple possible processing routes, allowing the job to choose its route during the manufacturing process.

Processing routes are allocated based on job types, resulting in $k$ unique pathways for different job processes. Let $\omega$ be the set of all processing routes in the system,  $\omega_{ik}=o_{i1k}o_{i2k} \dots o_{ijk}o_{i(j+1)k} \dots o_{ilk}$ be the $k$-th processing route of type-$b_i$, where $o_{ijk}$ represents the $j$-th operation on $\omega_{ik}$, and $l$ be the total number of operations required to complete the machining of the type-$b_i$ workpiece. The time required to process activity $o_{ijk}$ is denoted by $d_{ijk}$. 
{Each operation requires a specific types of resource, and two consecutive adjacent operations must be allocated different types of resources.}
For convenience, we add two operations $o_{S_i}$ and $o_{E_i}$ for each type of job that do not require any resource, which denote the storage of the original and finished machined parts for type-$b_i$ job, respectively. Thus, the routing with additional operations can be denoted as $\omega_{ik}=o_{S_i}o_{i1k}o_{i2k} \dots o_{ijk}o_{i(j+1)k} \dots o_{ilk}o_{E_i}$.

The processing route $\omega_{ik}$ of type-$b_i$ parts can be represented by a sequence of places and transitions $\rho(\omega_{ik}) = p_{S_i}t_{i11}p_{i11}t_{i12} \dots t_{ijk}p_{ijk}t_{i(j+1)k} \dots t_{ilk}p_{ilk}t_{E_i}p_{E_i}$, where $p_{ijk}$ is the operation place of the $k$-th processing route that executes the activity $o_{ijk}$, $t_{ijk}$ is a transition whose firing signifies the initiation of activity $o_{ijk}$, and the firing of $t_{E_i}$ represents the end of operation of type-$b_i$ parts.
The places $p_{S_i}$ and $p_{E_i}$ represent the start and end points of the machining path, i.e., the original part and the finished part, respectively.
We assign a resource place $p_i$ to resource $r_i$.
Note that different processing routes may involve some common resources, which means different operation-transition paths in the FMS might share identical resource places.
\begin{definition}
(\textit{Place-timed PN System for Scheduling}) A place-timed Petri net  system (P-TPNS) for FMS scheduling is a 7-tuple:
\[
\mathcal{N} = (P, T, Pre, Post, D, M_0, M_f),
\]
\begin{itemize}
    \item $P = P_M \cup P_S \cup P_{En} \cup P_R$ is the set of places, including operation (resp. resource) places set $P_M$ (resp. $P_R$), start (resp. end) places set $P_S$ (resp. $P_{En}$), where $P_S \nsubseteq P_M$, $P_{En} \nsubseteq P_M$ and $ P_R \cap (P_M \cup P_S \cup P_{En}) = \emptyset$.
    \item $T = T_S \cup T_{En}$ is the set of transitions, where $T_S$ and $T_{En}$ represent the sets of start transitions and end transitions, respectively. The subsets $T_S$ and $T_{En}$ are disjoint, i.e., $T_S \cap T_{En} = \emptyset$.
    \item $Pre: P \times T \rightarrow \{0, 1\}$ and $Post: T \times P \rightarrow \{0, 1\}$ are respectively \emph{pre-} and \emph{post-incidence function} which state the arcs connecting places and transitions.
    \item $D: P_M \to \mathbb{N}$ assigns time delays to operation places, with $D(p_{ijk}) = d_{ijk}$ for each place $p_{ijk} \in P_M$ and $D(p_{ijk}) = 0$ for each place $p_{ijk} \in P_S \cup P_{En} \cup P_R$.
    \item The initial marking $M_0$ is given by $M_0(p_{S_i}) = \varphi_i$ for each start place $p_{S_i} \in P_S$, $M_0(p_{ijk}) = 0$ for $p_{ijk} \in P_M \cup P_{En}$, and $M_0(p_i) = U(r_i)$ for each resource place $p_i \in P_R$.
    \item $M_f$ is a final marking, where $M_f(p_{E_i}) = \varphi_i$ for $p_{E_i} \in P_{En}$, $M_f(p_{ijk}) = 0$ for $p_{ijk} \in P_M \cup P_S$, and $M_f(p_i) = U(r_i)$ for $p_i \in P_R$. $\hfill\blacksquare$
\end{itemize}
\end{definition}

In Definition 1, $P_S$ and $P_{En}$ denote the storage of the original and finished machined parts, respectively, $T_S$ represents the start of the next operation as well as the end of the previous operation, and $T_{En}$ represents the completion of an job. For an operation place $p \in P_M$, the sets of input transitions $^{\bullet}{p}$ and output transitions ${p}^{\bullet}$ represent the start and end of the operation, respectively. The use and release of resources are modeled using directed arcs. If a resource place $p \in P_R$ needs to be used to perform an activity in an operation place $p_{ijk}$, then we have $Pre(p, t_{ijk}) = 1$ and $Post(p, t_{i(j+1)k}) = 1$. Initially, the number of tokens in the start place represents the number of raw materials for the activity, the presence of tokens in the operation place represents whether an operation is in progress, and the presence of tokens in the resource place represents whether a resource is available.

\begin{figure}[!htbp]
  \centering
  \includegraphics[scale=0.9]{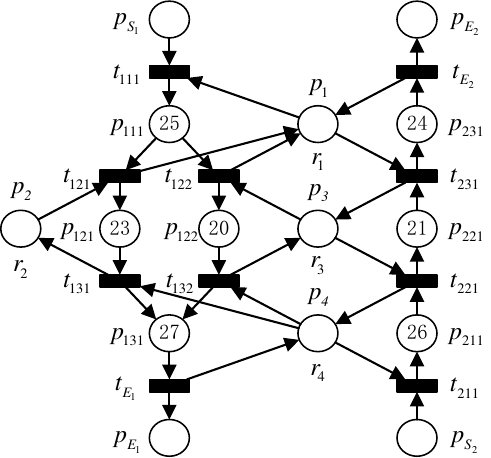}
  \caption{P-TPNS of the FMS in Example 1.}
  \label{example_fig1}
\end{figure}

\begin{example}
\label{example1}
We consider an FMS in \cite{Han2015} comprising four machines $R = \{r_1, r_2, r_3, r_4\}$. The jobs of the system requirements are outlined in Table \ref{table_1} with the operation times specified in parentheses. It is assume that each machine has a capacity of $U(r_i), r_i \in R$. The system can produce two types of jobs, i.e., $B = \{b_1, b_2\}$. The type-$b_1$ job follows two distinct processing routes, where the second operation can spend either 23 time units on $r_2$ or 20 time units on $r_3$. The first route is $\omega_{11} = o_{S_1}o_{111}o_{121}o_{131}o_{E_1}$, while the second route is $\omega_{12} = o_{S_1}o_{111}o_{122}o_{131}o_{E_1}$. The type-$b_2$ job follows a single processing route, $\omega_{21} = o_{S_2}o_{211}o_{221}o_{231}o_{E_2}$, moving in order through $r_4$, $r_3$, and $r_1$.

The P-TPNS of the FMS is shown in Fig.~\ref{example_fig1}. The processing route $\omega_{11}$ can be represented by $\rho(\omega_{11}) = p_{S_1}t_{111}p_{111}t_{121}p_{121}t_{131}p_{131}t_{E_1}p_{E_1}$. Similarly, we have $\rho(\omega_{12}) = p_{S_1}t_{111}p_{111}t_{122}p_{122}t_{132}p_{131}t_{E_1}p_{E_1}$ and $\rho(\omega_{21}) = p_{S_2}t_{211}p_{211}t_{221}p_{221}t_{231}p_{231}t_{E_2}p_{e_2}$. According to the definition, we have the set of start places $P_S=\{p_{S_1}, p_{S_2}\}$, the sets of end places $P_{En}=\{p_{E_1}, p_{E_2}\}$, the set of operation places $P_M=\{p_{111}, p_{121}, p_{122},p_{131},p_{211},p_{221}, p_{231}\}$, the set of resource places $ P_R=\{p_1, p_2, p_3, p_4\}$,  the set of  start transitions  $T_S=\{t_{111}, t_{121}, t_{131}, t_{122}, t_{132}, t_{211}, t_{221}, t_{231}\}$, the set of end transitions   $T_{En}=\{t_{En_1}, t_{En_2}\}$.~\hfill$\diamondsuit$ 
\end{example}

\begin{table}[!htbp]
\centering
\renewcommand{\arraystretch}{1.2}
\caption{Job requirements of the FMS in Example 1.}
\scalebox{1.2}{
\begin{tabular}{ccc}
\hline
 Job $b_{1}$        &Job $b_{2}$     \\ \hline
 $r_{1}$(25)    & $r_{4}$(26) \\
 $r_{2}$(23) or $r_{3}$(20) & $r_{3}$(21) \\
 $r_{4}$(27)    & $r_{1}$(24) \\ \hline
\end{tabular}
}
\label{table_1}
\end{table}

\begin{problem}
(\textit{Scheduling Problem}) Based on the P-TPNS, the scheduling problem can be described as finding a feasible transition sequence $\sigma^{\ast}$ from the initial marking $M_0$ to the final marking $M_f$ (i.e., $M_0[\sigma^{\ast} \rangle M_f$)  such that  the makespan $F_{max} = \max \{ \tau(o_{ilk}) \mid i = 1, 2, \cdots, s \}$ is minimized, 
where $\tau(o_{ilk})$ denotes the completion time of the last operation of job $b_i$, and $M_f$  denotes the  state where all jobs are completed and all resources are returned to their initial availability. 
\end{problem}

The objective of Problem~1 is to find a feasible transition sequence $\sigma^{\ast}$ that minimizes the makespan $F_{max}$, ensuring efficient utilization of resources and adherence to the precedence and timing constraints inherent in the P-TPNS. In this context, each valid transition sequence $\sigma^{\ast}$ corresponds to a feasible schedule. In particular, each transition $t_i$ in $\sigma^{\ast}$ represents a critical operation or an event in the FMS, such as the allocation or release of resources.

\subsection{Basis Marking and Basis Reachability Graph}
A compact structure of state space of a PN called \emph{basis reachability graph} (BRG) is proposed in \cite{Ma2017}, based on which the minimal sequence search problem can be solved efficiently. In the following, we recall some related concepts of BRG. For more details, we refer to \cite{Ma2017}, \cite{Ma2022}.

\begin{definition}
(\textit{Basis partition of transitions}) \cite{Ma2022} Given a PN $N = (P, T, Pre, Post)$, a basis partition of transition set $T$ is a pair $\pi = (T_E, T_I)$, where $T_E$ and $T_I$ are the sets of explicit transitions and implicit transitions, respectively. In particular, (i) $T_I \subseteq T$, $T_E = T \setminus T_I$; and (ii) the $T_I$-induced subnet is acyclic. $\hfill\blacksquare$
\end{definition}

\begin{definition}
(\textit{Explanations}) \cite{Ma2022} Given a PN $N$, a basis partition $\pi = (T_E, T_I)$, a reachable marking $M$, and a transition $t \in T_E$, we define:
\begin{itemize}
    \item $\Sigma(M, t) = \{\sigma \in T_I^* \mid M[\sigma\rangle M', M' \geq Pre(\cdot, t)\}$ as the set of \textit{explanations} of transition $t$ at marking $M$;
    \item $Y(M, t) = \{\mathbf{y}_{\sigma} \in \mathbb{N}^{n_I} \mid \sigma \in \Sigma(M, t)\}$ as the set of \textit{explanation vectors}, where $n_I =|T_I|$;
    \item $\Sigma_{min}(M, t) = \{\sigma \in \Sigma(M, t) \mid \nexists \sigma' \in \Sigma(M, t), \mathbf{y}_{\sigma'} \leq \mathbf{y}_{\sigma}\}$ as the set of\textit{ minimal explanations};
    \item $Y_{min}(M, t) = \{\mathbf{y}_{\sigma} \in \mathbb{N}^{n_I} \mid \sigma \in \Sigma_{min}(M, t)\}$ as the set of \emph{minimal explanation vectors}. $\hfill\blacksquare$
\end{itemize}
\end{definition}

Note that $\Sigma(M, t)$ denotes the set of implicit transition sequences that should fire at marking $M$ to enable an explicit transition $t$, $\Sigma_{min}(M, t)$ denotes the set of sequences in $\Sigma(M, t)$ with minimal transition sequences, and $Y_{min}(M, t)$ denotes the set of all minimal elements of $Y(M, t)$.

\begin{definition}
\label{BasisMarking}
(\textit{Basis marking}) \cite{Ma2022} Given a PN $N = (P$, $T$, $Pre$, $Post)$ with an initial marking $M_0$ and a basis partition $\pi = (T_E, T_I)$, the set of basis markings associated with $\pi = (T_E, T_I)$ is defined as
\begin{equation}
\begin{aligned}
\mathcal{M}(N, M_0, \pi) = \{\{M_0, M'\} \mid & M' = M_0 + C_I \cdot \mathbf{y} + C(\cdot, t), \\ & \mathbf{y} \in Y_{min}(M, t), t \in T_E\}.
\end{aligned}
\end{equation}
\end{definition}
A marking $M \in \mathcal{M}(N, M_0, \pi)$ is called a \textit{basis marking} under $\pi = (T_E, T_I)$ and $C_I$ denotes the incidence matrix $C$ restricted to $P \times T_I$. $\hfill\blacksquare$

\begin{definition}
\label{BRG}
(\textit{Basis Reachability Graph}) \cite{Ma2022} Given a PN $N = (P, T, Pre, Post)$ with an initial marking $M_0$ and a basis partition $\pi = (T_E, T_I)$, the basis reachability graph (BRG) associated with $\pi = (T_E, T_I)$ is a three-tuple $(\mathcal{M}(N, M_0, \pi), T_r, \Delta)$, where
\begin{itemize}
    \item the state set $\mathcal{M}(N, M_0, \pi)$ is the set of basis markings associated with $\pi = (T_E, T_I)$;
    \item the event set $T_r$ is the set of pairs $(t, \mathbf{y}) \in T_E \times \mathbb{N}^{n_I}$;
    \item $\Delta \subseteq \mathcal{M}(N, M_0, \pi) \times T_r \times \mathcal{M}(N, M_0, \pi)$ is a transition relation such that $\Delta = \{(M, (t, \mathbf{y}), M') \mid \mathbf{y} \in Y_{min}(M, t), t \in T_E, M' = M + C \cdot \mathbf{y} + C(\cdot, t)\}$. $\hfill\blacksquare$
\end{itemize}
\end{definition}

According to Definition \ref{BRG}, a PN has multiple BRGs depending on different basis partitions $\pi = (T_E, T_I)$. So far, there is no quantitative relationship between the size of the BRG and the choice of basis partition $\pi$. In this paper, we use the partitioning method in \cite{Ma2017} to obtain a good basis partition $\pi = (T_E, T_I)$ to reduce the size of the BRG as much as possible.

\begin{proposition}
\label{prop:1}
\cite{Ma2017} Given a P-TPNS and its BRG $\mathcal{B}=(\mathcal{M}(N, M_0, \pi), T_r, \Delta)$ associated with $\pi = (T_E, T_I)$, the following two statements are equivalent:
\begin{itemize}
    \item There exists a marking $M$ and a transition sequence $\sigma = \sigma_1 t_1 \cdots \sigma_n t_n \sigma_{n+1}$ ($\sigma_i \in T^*_I$ and $t_i \in T_E$) such that $M_0[\sigma\rangle M$;
    \item There exists a path $\mathcal{L}$ in the BRG defined as
    \[
    M_0 \xrightarrow{(t_1, \mathbf{y}_1)} M_{1} \xrightarrow{(t_2, \mathbf{y}_2)} \cdots \xrightarrow{(t_n, \mathbf{y}_n)} M_{n}
    \]
    such that $M \in \{M' \mid M_n[\sigma'\rangle M', \sigma' \in T^*_I \}$.
\end{itemize}
\end{proposition}


Typically, it is necessary to construct all the reachable markings of a P-TPNS to find a sequence $\sigma^{\ast}$ that satisfies the condition $M_0[\sigma^{\ast} \rangle M_f$, which not only imposes high memory requirements but also raises the scheduling complexity.
According to Proposition~\ref{prop:1}, it indicates that for any transition sequence $\sigma$ that consists of implicit transitions and explicit transitions in P-TPNS, we can equivalently find a transition sequence that is composed of only implicit transitions from the BRG.
Consequently, the BRG is much smaller compared to the reachability graph (RG), which reduces the storage space and improves the scheduling efficiency.
Although P-TPNS contains additional time constraints, we use its temporal information when performing state evaluation in the scheduling phase only, while we do not consider temporal information when constructing the BRG of P-TPNS. Thus the BRGs we construct for P-TPNS have the same basic structure as the BRGs for PNs.


\begin{example}
\label{example2}
(\textit{Example \ref{example1} continue}) We consider the P-TPNS in Fig. \ref{example_fig1} with an initial marking $M_0(p_{S_1}) = M_0(p_{S_2}) = 1$ and $M_0(p_i) = 1$, $i = 1, \dots, 4$, i.e., $M_0 = [1,0,0,0,0,0,1,0,0,0,0,1,1,1,1]^T$. 
The RG of the net has 26 reachable markings.
However, under a basis partition $\pi = (T_E, T_I)$ with $T_E = \{t_{121}, t_{122}, t_{E_1}, t_{221}, t_{E_2}\}$ and $T_I = T \setminus T_E$, the resulting BRG has only 11 basis markings, as shown in Fig. \ref{example_fig2}.
The minimal explanation vectors and basis markings of the BRG are shown in Table \ref{table_y} and Table \ref{table_2}, respectively.~\hfill$\diamondsuit$

\begin{figure}[!htbp]
  \centering
  \includegraphics[scale=0.65]{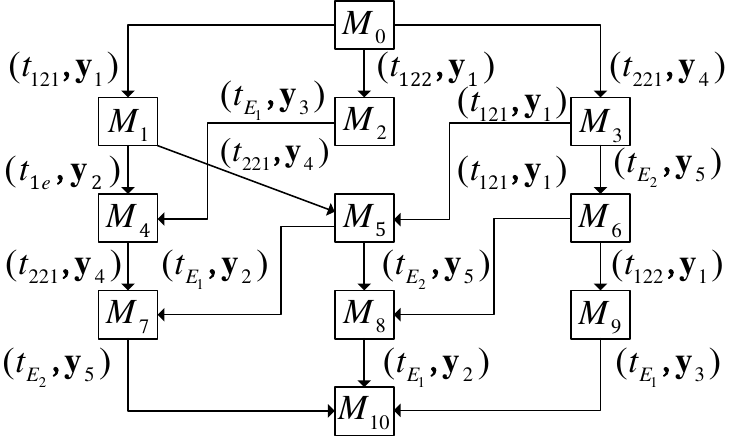}
  \caption{The BRG $\mathcal{B}$ of the P-TPNS in Fig. \ref{example_fig1}.}
  \label{example_fig2}
\end{figure}

\begin{table}[!htbp]
\centering
\renewcommand{\arraystretch}{1.2}
\caption{Minimal explanation vectors of Example \ref{example2}.}
\scalebox{1}{
\begin{tabular}{cccc}
\hline
Index          &   Minimal explanation vectors                         \\ \hline
$\mathbf{y}_1$ & [1, 0, 0, 0, 0, 0, 0, 0, 0, 0]\textsuperscript{T}    \\
$\mathbf{y}_2$ & [0, 0, 1, 0, 0, 0, 0, 0, 0, 0]\textsuperscript{T}    \\
$\mathbf{y}_3$ & [0, 0, 0, 0, 1, 0, 0, 0, 0, 0]\textsuperscript{T}    \\
$\mathbf{y}_4$ & [0, 0, 0, 0, 0, 0, 1, 0, 0, 0]\textsuperscript{T}    \\
$\mathbf{y}_5$ & [0, 0, 0, 0, 0, 0, 0, 0, 1, 0]\textsuperscript{T}    \\ \hline
\end{tabular}
}
\label{table_y}
\end{table}

\begin{table}[!htbp]
\centering
\renewcommand{\arraystretch}{1.2}
\caption{Basis markings in the BRG $\mathcal{B}$ of Example \ref{example2}.}
\scalebox{0.85}{
\begin{tabular}{cccc}
\hline
Index & Basis markings    & Index & Basis markings  \\ \hline
$M_0$ & [1,0,0,0,0,0,1,0,0,0,0,1,1,1,1]\textsuperscript{T}    & $M_6$ & [1,0,0,0,0,0,0,0,0,0,1,1,1,1,1]\textsuperscript{T} \\
$M_1$ & [0,0,1,0,0,0,1,0,0,0,0,1,0,1,1]\textsuperscript{T}    & $M_7$ &
[0,0,0,0,0,1,0,0,1,0,0,1,1,0,1]\textsuperscript{T} \\
$M_2$ & [0,0,0,1,0,0,1,0,0,0,0,1,1,0,1]\textsuperscript{T}    & $M_8$ &
[0,0,1,0,0,0,0,0,0,0,1,1,0,1,1]\textsuperscript{T} \\
$M_3$ & [1,0,0,0,0,0,0,0,1,0,0,1,1,0,1]\textsuperscript{T}    & $M_9$ &
[0,0,0,1,0,0,0,0,0,0,1,1,1,0,1]\textsuperscript{T} \\
$M_4$ & [0,0,0,0,0,1,1,0,0,0,0,1,1,1,1]\textsuperscript{T}    & $M_{10}$ &
[0,0,0,0,0,1,0,0,0,0,1,1,1,1,1]\textsuperscript{T} \\
$M_5$ & [0,0,1,0,0,0,0,0,1,0,0,1,0,0,1]\textsuperscript{T}    &  & \\ \hline
\end{tabular}
}
\label{table_2}
\end{table}
\end{example}

\section{Generation filtered beam search  method for FMS scheduling}\label{4}
Heuristic search algorithm expands the most promising branches of a search tree according to a criterion established with the heuristic function. In particular, 
{beam search holds an advantage over the A$^{\ast}$ algorithm primarily due to its reduced memory usage and improved computational efficiency. While A$^{\ast}$ retains all possible nodes, leading to greater memory consumption and increased time complexity, beam search addresses these issues by restricting the number of candidate nodes at each expansion, governed by the ``beam width".}  In this section, an improved beam search algorithm is proposed to solve the scheduling problem of FMSs based on the constructed BRG, where the algorithm extends the BRG by combining transition sequences.
It can accelerate the scheduling process and alleviate the problem of low scheduling efficiency based on the RG and excessive growth of scheduling time with the expansion of the system size.

\subsection{States of a P-TPNS}
For a basis marking $M \in \mathcal{M}(N, M_0, \pi)$ in the BRG $\mathcal{B}$, there exist multiple feasible transition sequences capable of driving the net from the initial marking $M_0$ to $M_f$. This implies that the same marking can be obtained through various routes, which can lead to differences in completion time. In order to use the information from transition sequences in heuristic searches within the BRG, this paper extends the BRG by associating each reachable marking $M$ with its corresponding transition sequence $\sigma$.

The state of a P-TPNS $\mathcal{N}$ is defined as a pair $V = (M, \sigma)$, where $M \in \mathcal{M}(N, M_0, \pi)$ denotes the basis marking in $\mathcal{B}$ and $\sigma$ denotes its corresponding feasible transition sequence, such that $M_0[\sigma \rangle M$. It is obvious that $V$ is similarly affected by different basis partitions $\pi = (T_E, T_I)$. $V_0 = (M_0, \varepsilon)$ denotes an initial state of P-TPN. Simply put, a state includes both the marking and the transition sequence that leads to that marking. Two states $(M_1, \sigma_1)$ and $(M_2, \sigma_2)$ are considered identical if $M_1 = M_2$ and $\sigma_1 = \sigma_2$. 

\subsection{Real cost of a state}
The scheduling problem addressed in this paper involves determining a feasible schedule to process a given set of parts in an FMS to minimize the makespan. A novel heuristic function is introduced based on the state $V$. The fitness function takes the form $f(M, \sigma) = g(M, \sigma) + h(M, \sigma)$, where $g(M, \sigma)$ is the completion time related to the transition sequence $\sigma$, and $h(M, \sigma)$ is derived from the weighted operation time \cite{Huang2022}. The term $g(M, \sigma)$ is the real cost from the initial state to a current state $(M, \sigma)$ and $h(M, \sigma)$ is an estimate of the cost from such a state to a desired goal state.

\begin{definition}
(\textit{Trajectory of a state}) Given a state $V=(M_i, \sigma)$ with $\sigma = \sigma_1 t_1 \dots \sigma_i t_i$,  the  trajectory of state $V$ is a sequence $M_0[\sigma_1 t_1 \rangle M_1 \dots [\sigma_{i-1} t_{i-1} \rangle M_{i-1}[\sigma_i t_i \rangle M_i$, where $\sigma_i$ is the minimal explanation of $t_i$ at  basis marking $M_{i-1}\in \mathcal{M}(N, M_0, \pi)$. $\hfill\blacksquare$
\end{definition}

\begin{definition}
(\textit{Set of processing times}) The set of processing times on resource $r_z$ is defined as
\begin{equation}
\begin{aligned}
\label{eq5}
X(r_z) = \{[\chi_{ijS}, \chi_{ijE}) \mid  i \in\{1,\ldots,q\},& j \in \{1,\ldots,l\}, \\ 
& z \in \{1,\ldots,s\}\}
\end{aligned}
\end{equation}
in which $\chi_{ijS}$ and $\chi_{ijE}$ denote the start time and end time of the $j$-th process of type-$b_i$ job, respectively. $\hfill\blacksquare$
\end{definition}

The capacity $U(r_z)$ of the resource $r_z$ denotes the maximum number of jobs that can be produced at the same time. Then the processing times of the jobs on the resource $r_z$ can overlap.

\begin{definition}
(\textit{Resource utilization}) The number of jobs being machined on resource $r_z$ at time $\chi$ is
\begin{equation}
\label{eq6}
L(X(r_z), \chi) \leqslant U(r_z), \forall \chi \in [\chi_{ijS}, \chi_{ijE})
\end{equation}
which cannot exceed the total number of capacity in resource $U(r_z)$ to avoid resource overload. $\hfill\blacksquare$
\end{definition}

\begin{definition}
(\textit{Idle time interval}) The idle time interval refers to the period when the upper limit of the number of machines is not reached at a given moment $\chi$.
The idle time interval of resource $r_z$ is defined as
\begin{equation}
\label{eq7}
\begin{aligned}
I(r_z) = \{[\chi_s, \chi_e) \mid \forall \chi \in [\chi_s, \chi_e), L(X(r_z), \chi) < U(r_z)\}.
\end{aligned}
\end{equation}
where $\chi_s$ and $\chi_e$ denote the start and end of idle time, respectively. $\hfill\blacksquare$
\end{definition}

An operation $o_{ijk}$ can be scheduled during each idle time period $[\chi_s, \chi_e)$. Its processing time period $[\chi_{ijS}, \chi_{ijE})$ should be scheduled in idle time interval of the resource, i.e., $[\chi_{ijS}, \chi_{ijE}) \subseteq I(r_z)$. When scheduling the processing of operations, it should be ensured that $L(X(r_z), \chi) < U(r_z)$, i.e., at any moment $\chi$, the number of operations on resource $r_z$ does not exceed the capacity limit of resource $U(r_z)$. In addition, during the scheduling process, it is essential to allocate new workpieces for machining within the {\em idle time intervals} of a machine, which represent the available machining slots.

\begin{algorithm}
\caption{Find Idle Time Interval}
\label{algorithm1}
\KwIn{processing time $X(r_z)$ on $r_z$, capacity $U(r_z)$ of resource $r_z$}
\KwOut{Idle time intervals $I(r_z)$}

\SetKwFunction{FindIdleTime}{Find\_Idle\_Time}
\SetKwProg{Fn}{Function}{:}{}

\Fn{Find\_Idle\_Time($X(r_z)$, $U(r_z)$)}{
    Initialization: $I(r_z) \leftarrow \{\}$; $W \leftarrow 0$, $\chi_s \leftarrow 0$, $\chi_e \leftarrow 0$\;
    \For{each $[\chi_{ijS}, \chi_{ijE})$ in $X(r_z)$}{
        Tag $\chi_{ijS}$ with ``start'';
        Tag $\chi_{ijE}$ with ``end'';
    }
    Sort $X(r_z)$ by time, with ``end'' events preceding ``start'' events at the same time\;

    \For{each $\chi$ in $X(r_z)$}{
        \If{Tag of $\chi$ is ``start''}{
            Increment $W$\;
            \If{$W = U(r_z)$}{
                Set $\chi_s \gets \chi$\;
            }
        }
        \ElseIf{Tag of $\chi$ is ``end''}{
            \If{$W = U(r_z)$ and $\chi_s \neq \chi$}{
                $\chi_e \gets \chi$\;
                Append interval $[\chi_s, \chi_e)$ to $I(r_z)$\;
            }
            Decrement $W$\;
        }
    }

    Set $I(r_z) \leftarrow [0, +\infty) \setminus I(r_z)$\;
    \KwRet{$I(r_z)$}\;
}

\end{algorithm}

Computation of the idle time interval of resource $r_z$ is shown in Algorithm \ref{algorithm1}. The algorithm determines the processing times when a resource reaches its maximum load by tagging events and sequencing processing times of the machine, and then it calculates idle time of the resource by subtracting that time from the entire timeline.

\begin{example}[Example \ref{example2} Continued]
\label{example3}
For the P-TPN with $M_0(p_{S_1}) = M_0(p_{S_2}) = 2$ and $M_0(p_i) = 2$ ($i = 1, \dots, 4$) in Fig. \ref{example_fig1}, the initial marking is $M_0 = [2,0,0,0,0,0,2,0,0,0,0,2,2,2,2]^T$. Marking $M_1 = [1,0,0,0,1,0,1,0,1,0,0,2,2,0,2]^T$ is obtained from firing events $(t_{122}, \mathbf{y_1})$ and $(t_{211}, \mathbf{y_4})$. The processing times on the resources are $X(r_1) = \{[0, 25)\}$, $X(r_3) = \{[25, 45), [26, 47)\}$, and $X(r_4) = \{[0, 26)\}$, respectively. There are no processing operations on resource $r_2$. According to Algorithm \ref{algorithm1}, the start time $\chi_{ijS}$ and the end time $\chi_{ijE}$ of each processing interval are first labeled as ``start'' and ``end'', respectively. The $X(r_3)$ becomes $\{(25, \text{``start''}), (45, \text{``end''}), (26, \text{``start''}), (47, \text{``end''})\}$. Next, $X(r_3)$ is sorted chronologically, ensuring that the ``end'' event is prioritized over the ``start'' event when occurring at the same time. The sorted $X(r_3)$ is $\{(25, \text{``start''}), (26, \text{``start''}), (45, \text{``end''}), (47, \text{``end''})\}$.

During the traversal of $X(r_3)$, when a ``start'' event is encountered, the counter $W$ increases. If $W$ equals $U(r_3)$, the corresponding time is labeled as the start of the overlap interval. When traversing to the start time 26, we have $W = U(r_3) = 2$ and $\chi_s = 26$. When an ``end'' event is encountered, the counter $W$ is decremented. If counter $W$ equals $U(r_3)$ and $\chi_s$ has been recorded, the end time $\chi_e$ is marked and the interval $[\chi_s, \chi_e)$ is added to $I(r_3)$. The next time is 45 with an event ``end'', i.e., $\chi_e = 45$. Eventually, $I(r_3) = [26, 45)$ holds. By removing these time intervals from the timeline, the resulting free time intervals are $I(r_3) = \{[0, 26), [45, +\infty)\}$, representing the available free time periods for new tasks on resource $r_3$.~\hfill$\diamondsuit$
\end{example}

After determining the idle time of each resource, when a new operation needs to be scheduled, it is reasonably scheduled to be processed within the idle time interval of the corresponding resource based on its required processing time and provided that the relevant processing constraints are met. This scheduling method ensures the effective utilization of resources and avoids machine overload or delays.

The calculation of the real cost $g(M, \sigma)$ of a state $(M, \sigma)$ is shown in Algorithm \ref{algorithm2}. Note that we use as input to Algorithm \ref{algorithm2} the transition sequences $\sigma_n t_n$ corresponding to the last event $(t_n, \mathbf{y}_n)$ of $\sigma$ that can produce a basis marking $M$. Briefly, each process in the transition sequence is processed in turn, determining the resource it uses and calculating its idle time interval on the current resource. For each process, a start time that matches the idle interval is selected if it is the first scheduling; if it is a subsequent process scheduling, an adjustment is made based on the end time of the previous process. Finally, the maximum end time $g(M, \sigma)$ of a task is found by traversing the processing time interval $X(r_z)$ of all tasks.

\begin{example}[Example \ref{example3} Continued]
\label{example4}
Consider the event $(t_{E_1}, \mathbf{y_3})$ at $(M_1, \sigma_1)$ that generates a new state $(M_2, \sigma_2)$, where $\mathbf{y_3} = [0, 0, 0, 0, 1, 0, 0, 0, 0, 0]^T$ and $M_2 = [1,0,0,0,0,1,1,0,1,0,0,2,2,1,2]^T$. The transition sequence corresponding to $(t_{E_1}, \mathbf{y_3})$ is $t_{132}t_{E_1}$ and $\sigma_2 = \sigma_1t_{132}t_{E_1}$. According to Algorithm \ref{algorithm2}, the resource $r_4$ is used  for the output places $p_{131}$ of $t_{132}$ and we can use Algorithm \ref{algorithm1} to compute $I(r_4) = \{[0, +\infty)\}$. Since the start processing time of the previous operation is not empty, i.e., $\chi_{12S} = 45$, we select $\chi_{13S} = 45 \ge \max\{\chi_{12S}\}$ and $\chi_{13E} = \chi_{13S} + D(p_{131}) = 72$ such that it satisfies $[\chi_{13S}, \chi_{13S} + D(p_{131})) \subseteq I(r_4)$. Then $X(r_4) \cup [\chi_{13S}, \chi_{13E}) = \{[0, 26), [45, 72)\}$ holds. Since there is no resource used for the output places of $t_{E_1}$, no operations need to be done. Finally, the algorithm outputs $g(M_2, \sigma_2) = max\{25, 26, 45, 47, 72\} = 72$.~\hfill$\diamondsuit$
\end{example}

\begin{algorithm}
\caption{Computation of  the Real Cost $g(M, \sigma)$}
\label{algorithm2}
\KwIn{Set of processing times $X$ on $r_z$, $\sigma_n t_n$}
\KwOut{The Real Cost $g(M, \sigma)$}

\SetKwFunction{ComputeRealCost}{Compute\_Real\_Cost}
\SetKwProg{Fn}{Function}{:}{}

\Fn{Compute\_Real\_Cost($X$, $\sigma_n t_n$)}{
    \For{each $t_{ijk} \in \sigma_n t_n$}{
        Select the resource $r_z$ used by the output places $p$ of $t_{ijk}$\;
        $I(r_z) \leftarrow Find\_Idle\_Time\{X(r_z), U(r_z)\}$\;
        \If{start processing time of the previous operation $\chi_{i(j-1)S}$ is empty}{ 
            Select $[\chi_{ijS}, \chi_{ijS} + D(p)) \subseteq I(r_z)$\;
            $\chi_{ijE} \gets \chi_{ijS} + D(p)$\;
            $X(r_z) \cup [\chi_{ijS}, \chi_{ijE})$\;
        }
        \ElseIf{start processing time of the previous operation $\chi_{i(j-1)S}$ is not empty}{ 
            Select $\chi_{ijS} \ge \max\{\chi_{i(j-1)E}\}$ and $[\chi_{ijS}, \chi_{ijS} + D_(p) \subseteq I(r_z)$\;
            $\chi_{ijE} \gets \chi_{ijS} + D(p)$\;
            $X(r_z) \cup [\chi_{ijS}, \chi_{ijE})$\;
        }
    }

    $g(M, \sigma) \gets max\{\chi_{ijE}\}$, where $\chi_{ijE} \in X$\;

    \KwRet{$g(M, \sigma)$}\;
}

\end{algorithm}

\subsection{Heuristic Function}
During the heuristic search, not only the real cost of a state needs to be evaluated using Algorithm \ref{algorithm2}, but also the estimated cost from the current state to the final state, as determined by the heuristic function.
In this paper, we adpot the heuristic function from \cite{Huang2022}, where two matrices are proposed to compute a lower bound on the cost from an initial state to an final state: a weighted operation time (WOT) matrix $\Theta$ and a weighted resource time (WRT) matrix $\Gamma$. The dimension of matrix $\Theta$ is $|P_M| \times |P_R|$, where for any operation place $p$ that requires a type of resource $r$, the entry $\Theta(p, r)$ represents the minimul operation time needed by an available token in $ p $, assuming all units of $ r $ can be used at the same time. The dimension of matrix $\Gamma$ is $|P \setminus P_R| \times |P_R|$, where for each $p \in P \setminus P_R$ and $r \in P_R$, the entry $\Gamma(p, r)$ represents the minimal cumulative WOT required for an available token at place $p$ to reach its terminal place through the resource $r$.
The heuristic function for scheduling the FMS can be given as
\begin{equation}
\label{eq8}
h(M, \sigma)=\max_{r\in P_{R}}\{\sum_{p\in P\setminus P_{R}}[M(p)\cdot\Gamma(p,r)]\}.
\end{equation}

\subsection{Generation Filtered Beam Search Algorithm}
{In this section, based on the expansion of the states of a BRG, we develop a generation filtered beam search (GFBS) \cite{Cherif2019} algorithm that is} an improved version of the traditional filtered beam search algorithm to solve Problem~1 effciently. By sorting the generated candidate states, the algorithm ensures that each generation of candidate states has a chance to be selected, thus avoiding the local optimal solution problem caused by premature elimination of potentially high-quality candidates.

The main innovations of the GFBS are as follows. Firstly, it introduces a sub-generation mechanism to comprehensively expand and sort all candidate states of the same generation instead of only locally expanding the optimal parent states; secondly, the algorithm uses a heuristic cost function that combines the actual time of the computed trajectories and the estimation time of the unfinished trajectories, which effectively reduces the computational complexity in the search process. The algorithm gradually approximates target markers by iteratively generating candidate states, thus finding the optimal or near-optimal scheduling path.

The GFBS algorithm maintains the $OPEN$, $TEMPLIST$, and $GENERATION$ lists to store the current generation of parent states, the successor states generated by each parent state, and the optimal to-be-extended successor states, respectively. It involves two critical parameters: the global parameter $\beta_g$ and the local parameter $\beta_l$. The global parameter $\beta_g$ determines the number of top parent states from the $GENERATION$ list which are retained in the $OPEN$ list for further exploration during each iteration. In contrast, the local parameter $\beta_l$ specifies the number of top-level successor states to be selected from each parent state during the generation of candidate states, i.e., at most $\beta_l$ successor states from each parent state can be retained in the $GENERATION$ list. Meanwhile, these parameters control the trade-off between exploration diversity and computational efficiency. A larger $\beta_g$ and $\beta_l$ enhances the algorithm's ability to explore the state space comprehensively, potentially leading to higher-quality solutions. However, it also increases the computational burden. Conversely, smaller values reduce computational costs, but may increase the risk of converging to suboptimal solutions.

To utilize the information of transition sequences, we use the GFBS algorithm to search the path from $M_0$ to $M_f$.
The algorithm aims to return a valid transition sequence $\sigma$ and the makespan $F_{max}$.
The main innovation of the proposed algorithm is based on the basis partitions of the delineated transitions, the basis markings of the P-TPNS are gradually explored until the final marking is found. 
Note that during the search process, multiple explorations may lead to the same basis marking, while a single basis marking can correspond to various distinct transition sequences.

\begin{algorithm}
\caption{GFBS Algorithm for FMS scheduling}
\label{algorithm3}
\KwIn{$\mathcal{N}$, $(M_0, \varepsilon)$, $\beta_g$, $\beta_l$}
\KwOut{$\sigma$, $F_{max}$}
Initialization: $OPEN$$\leftarrow \{\}$, $TEMPLIST$$\leftarrow \{\}$, $GENERATION$$\leftarrow \{\}$\;
Place $(M_0, \varepsilon)$ into $OPEN$\;

\While{OPEN is not empty}{
    \For{each candidate $(M, \sigma)$ remaining in $OPEN$}{
        \If{$M = M_{f}$}{
                \Return{$\sigma$ and $F_{max} = g(M, \sigma)$}\;
            }
        \For{each $t \in T_E$}{
            \For{each $\mathbf{y} \in Y_{min}(M, t)$}{
                $M' = M + C \cdot \mathbf{y} + C(\cdot, t)$;
                $\sigma' = \sigma \cup (t, \mathbf{y})$;

                Calculate the cost function $f(M', \sigma') = g(M', \sigma') + h(M', \sigma')$\;
                Place $(M', \sigma')$ into a temporary list $TEMPLIST$\;
            }
        }
        Sort $TEMPLIST$ in ascending order of $f(M', \sigma')$\ and keep up to $\beta_l$ best states;

        \For{each state $(M', \sigma')$ in $TEMPLIST$}{
            \If{$GENERATION$ contains a state $(M_G, \sigma_G)$ identical to $(M', \sigma')$}{
                \If{$g(M', \sigma') < g(M_G, \sigma_G)$}{
                    Replace $(M_G, \sigma_G)$ with $(M', \sigma')$ in $GENERATION$\;
                }
            }
            \Else{
                place $(M', \sigma')$ to $GENERATION$\;
            }
        }

        Clear $TEMPLIST$\;
    }
    Clear $OPEN$\;
    Select up to $\beta_g$ best states from $GENERATION$ and place them into $OPEN$\;
    Clear $GENERATION$\;
}
\Return{$\sigma= \varepsilon$ and $F_{max} = \infty$ if no solution is found}\;
\end{algorithm}

\begin{figure*}[!htbp]
  \centering
  \includegraphics[scale=0.52]{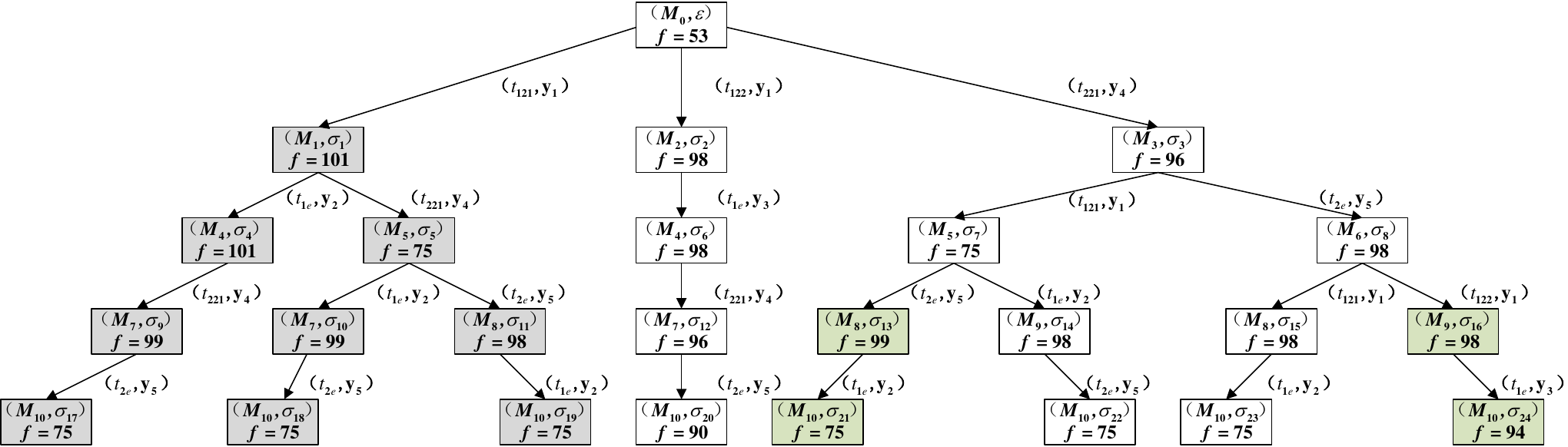}
  \caption{Schematic diagram of GFBS algorithm search for P-TPNS in Example \ref{example2}.}
  \label{GFBSexample}
\end{figure*}

The developed GFBS algorithm is outlined in Algorithm~\ref{algorithm3}. The algorithm leverages the BRG and incorporates transition sequences into state comparison, ensuring a robust and efficient exploration of the solution space. By associating each reachable basis mark $M$ with its corresponding transition sequence $\sigma$, the algorithm effectively differentiates states with identical markings but varying transition histories, improving the quality of the solution. The algorithm begins by initializing the $OPEN$  with a state $(M, \sigma)$ that consists of the initial marking $M_0$ and an empty transition sequence $\varepsilon$, while $TEMPLIST$ and $GENERATION$  are used to store intermediate states.
At each iteration, all candidate states in $OPEN$ are expanded to generate successors, and their fitness $f(M', \sigma') = g(M', \sigma') + h(M', \sigma')$ are computed.
The algorithm selects the top $\beta_l$ successors and stores them in $GENERATION$ based on the fitness, where duplicate states with different transition sequences are resolved by retaining the one with smaller real cost $g(M', \sigma')$.
After processing, the top $\beta_g$ candidates from $GENERATION$ are put back into $OPEN$ for the next iteration.
The process continues until the final marking $M_f$ is reached, returning the feasible transition sequence $\sigma$ and makespan $F_{max}$, or it terminates with no solution if the search space is exhausted.

\begin{example}[Example \ref{example2} Continued]
\label{example5}
Consider again the P-TPNS shown in Fig. \ref{example_fig1} with an initial marking $M_0 = [1,0,0,0,0,0,1,0,0,0,0,1,1,1,1]^T$ and a basis partition $\pi = (T_E, T_I)$, where $T_E = \{t_{121}, t_{122}, t_{E_1}, t_{221}, t_{E_2}\}$ and $T_I = T \setminus T_E$. The search schematic of the GFBS algorithm with $\beta_l = 2$ and $\beta_g = 3$ is shown in Fig. \ref{GFBSexample}. The gray color in Fig. \ref{GFBSexample} represents the states filtered out by $\beta_l$, and the light green color represents the states filtered out by $\beta_g$. The algorithm firstly adds the initial state $(M_0, \varepsilon)$ to $OPEN$. If $OPEN$ is not empty, the states are taken sequentially to calculate their reachable states.
The real cost and heuristic function values for each state are evaluated according to Algorithm \ref{algorithm2} and (\ref{eq8}), i.e., $(M_1, \sigma_1), (M_2, \sigma_2), (M_3, \sigma_3)$ and $f(M_1, \sigma_1) = g(M_1, \sigma_1) + h(M_1, \sigma_1) = 101, f(M_2, \sigma_2) = g(M_2, \sigma_2) + h(M_2, \sigma_2) = 98, f(M_3, \sigma_3) = g(M_3, \sigma_3) + h(M_3, \sigma_3) = 96$.
After putting them into $TEMPLIST$, it holds $TEMPLIST$ $=[(M_1, \sigma_1), (M_2, \sigma_2), (M_3, \sigma_3)]$.
We sort $TEMPLIST$ according to $f(M, \sigma)$ and then add the first $\beta_l$ states to $GENERATION$ if there is the same state then judge whether to replace it according to the size of $g(M, \sigma)$.
In this sense, $GENERATION=[(M_2, \sigma_2), (M_3, \sigma_3)]$ holds.
Then we clear $TEMPLIST$ when all the states in $OPEN$ are searched, and add the first $\beta_g$ states of $GENERATION$ to $OPEN$ and continue to complete the next round of search until the search reaches the termination state.~\hfill$\diamondsuit$
\end{example}

The algorithm effectively reduces the computational burden by leveraging the compact representation of RG and dynamically pruning the search space. The time complexity of each iteration depends on the beam widths $\beta_g$ and $\beta_l$, as well as the size of the BRG.
The search process of the algorithm when $\beta_l = \beta_g = \infty$ is shown in Fig. \ref{GFBSexample}, where the algorithm does not filter the searched states. In such a situation, both the computational complexity and the accuracy of the solution are high.
The algorithm search process with $\beta_l = 2$ and $\beta_g = 3$ is shown in the white part of Fig. \ref{GFBSexample}, where gray parts are the states filtered out by $\beta_l$ and green parts are the states filtered out by $\beta_g$.
When $\beta_l = \beta_g= 1$ the algorithm retains only one state per layer so its complexity is low but accuracy is poor, its searched states per layer in the graph are $(M_0, \varepsilon), (M_3, \sigma_3), (M_5, \sigma_7), (M_9, \sigma_{14}), (M_{10}, \sigma_{22})$ in that order.

\section{Experimental Results}\label{5}
The proposed algorithm in this paper is coded in Python and runs on a 2.1 GHz personal computer with 16G RAM memory. 
In the following subsections, a series of benchmark systems are tested and compared with some existing approaches.  The source code for all the tested examples is available online\footnote{https://github.com/ZhouHe0052/FMS-Scheduling}.

\subsection{Illustrative Example}
In this subsection, the GFBS algorithm is performed on the FMS taken from \cite{Huang2022In1}. The system consists of two robots (denoted as $r_1$ and $r_2$), four types of machines (denoted by $r_3,\ldots, r_6$), each having one unit available. Furthermore,  machines and robots process two categories of jobs or parts, identified as $b_1$ and $b_2$. The processing times and resources required by the operations are given in Table \ref{table_3}, with the operation time for each activity shown in parentheses.

\begin{table}[!htbp]
\centering
\renewcommand{\arraystretch}{1.2}
\caption{Process routes of the FMS in Fig \ref{example_fig3}.}
\scalebox{1.2}{
\begin{tabular}{ccc}
\hline
   & $b_{1}$    & $b_{2}$ \\ \hline
$o_{1}$ & $r_{1}$(3)    & $r_{2}$(2) \\
$o_{2}$ & $r_{1}$(4) or $r_{4}$(2) & $r_{6}$(4) \\
$o_{3}$ & $r_{1}$(4)    & $r_{1}$(4) \\
$o_{4}$ & $r_{5}$(3)    & $r_{4}$(3) \\
$o_{5}$ & $r_{2}$(5)    & $r_{1}$(5) \\ \hline
\end{tabular}
}
\label{table_3}
\end{table}

\begin{figure}[!htbp]
  \centering
  \includegraphics[scale=0.6]{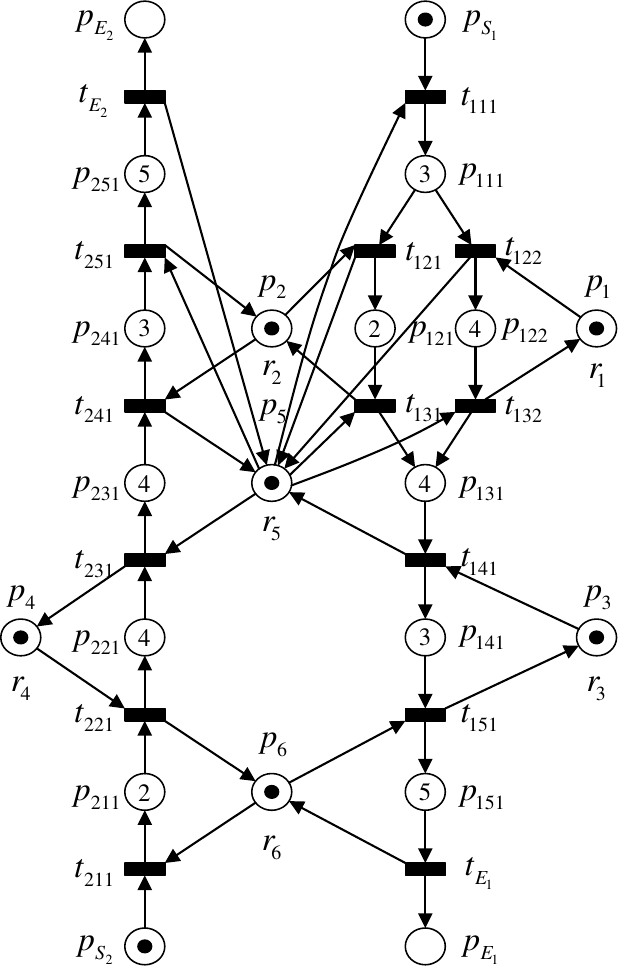}
  \caption{The P-TPNS of an FMS taken from \cite{Huang2022In1}.}
  \label{example_fig3}
\end{figure}

\begin{figure}[!htbp]
  \centering
  \includegraphics[scale=0.7]{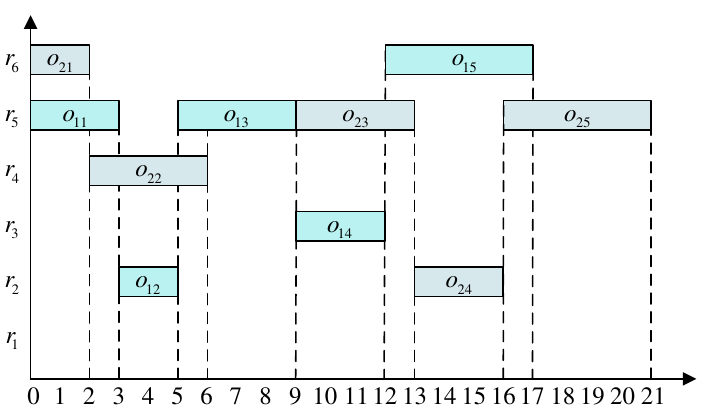}
  \caption{Gantt chart of schedule $Seq$.}
  \label{Gantt_fig4}
\end{figure}

 The P-TPNS of the FMS with 14 transitions and 21 places is shown in Fig. \ref{example_fig3}. The capacities of the jobs and resources are both equal to one, i.e., $M_0(p_{S_i}) = M_0(p_j) = 1$, for $i = 1, 2$, $j = 1, \dots, 6$.
In the final marking, it holds $M_f(p_{E_i}) = M_f(p_j) = 1$ for $i = 1, 2$, $j = 1, \dots, 6$.
Moreover, the sets of explicit transitions and implicit transitions are  $T_E = \{t_{121}, t_{122}, t_{141}, t_{E_1}, t_{221}, t_{241}, t_{E_2}\}$ and $T_I = \{t_{111}, t_{131}, t_{132}, t_{151}, t_{211}, t_{231}, t_{251}\}$, respectively.
The algorithm returns a makespan $F_{max} = 21$ and a transition sequence $Seq = \{t_{111}, t_{121}, t_{211}, t_{221}, t_{131}, t_{141}, t_{231}, t_{241}, t_{151}, t_{E_1}, t_{251}, t_{E_2}\}$ assuming that $\beta_g = 3$ and $\beta_l = 2$.
The Gantt chart of the sequence $Seq$ is shown in Fig. \ref{Gantt_fig4}.
In addition, the GFBS algorithm with comparison of transition sequences can obtain the same $F_{max}$ as \cite{Huang2022In1} in $1$s.

 In addition, the scheduling results and the running times for different basis partitions are presented in Table \ref{table_4}, where only the implicit transition set $T_I$ is shown (the explicit transition set $T_E$ can be obtained by $T_E = T \setminus T_I$). It can be seen that the GFBS algorithm obtains the optimal solutions $F_{max} = 21$ in all cases with different basis partitions.
Moreover, the difference in the running time of the algorithm is quite small. As a consequence, the division of the basis partition does not affect the GFBS algorithm too much.


\begin{table}[!htbp]
\centering
\renewcommand{\arraystretch}{1.2}
\caption{Scheduling results for different basis partitions.}
\scalebox{1.2}{
\begin{tabular}{ccc}
\hline
$T_I$  & $F_{max}$    & CPU(s) \\ \hline
$t_{111}, t_{141}, t_{211}, t_{241}$                            & 21    & 0.015s \\
$t_{121}, t_{132}, t_{151}, t_{221}, t_{241}$                   & 21    & 0.014s \\
$t_{121}, t_{122}, t_{151}, t_{211}, t_{231}$                   & 21    & 0.012s \\
$t_{121}, t_{132}, t_{151}, t_{211}, t_{231}$                   & 21    & 0.012s \\
$t_{122}, t_{131}, t_{151}, t_{211}, t_{231}, t_{251}$          & 21    & 0.012s \\
$t_{111}, t_{131}, t_{132}, t_{151}, t_{211}, t_{231}, t_{251}$ & 21    & 0.011s \\ \hline
\end{tabular}
}
\label{table_4}
\end{table}
\begin{table*}[!htbp]
\centering
\renewcommand{\arraystretch}{1.2}
\caption{Scheduling results for Example 2 in \cite{Huang2022}.}
\scalebox{1.2}{
\begin{tabular}{lllcrr|crr|crr}
\hline
\multicolumn{1}{r}{\multirow{2}{*}{$b_1$}} & \multirow{2}{*}{$b_2$} & \multirow{2}{*}{$b_3$}
& \multicolumn{3}{c|}{GFBS} & \multicolumn{3}{c|}{IHFBS\cite{Meja2017}} & \multicolumn{3}{c}{A$^{\ast}$\cite{Huang2022}} \\ \cline{4-12}
\multicolumn{1}{r}{} & &
& \multicolumn{1}{c}{$F_{max}$} & \multicolumn{1}{r}{$\mathcal{N}$} & \multicolumn{1}{r|}{CPU[s]}
& \multicolumn{1}{c}{$F_{max}$} & \multicolumn{1}{r}{$\mathcal{N}$} & \multicolumn{1}{r|}{CPU[s]}
& \multicolumn{1}{c}{$F_{max}$} & \multicolumn{1}{r}{$\mathcal{N}$} & \multicolumn{1}{r}{CPU[s]} \\ \hline
1 & 1 & 1 & \textbf{21}  & 117  & 0.01 & \textbf{21}  & 87069 & 33.00  & \textbf{21} & 477   & 0.39       \\
2 & 2 & 2 & \textbf{30}  & 368  & 0.08 & 32           & 126496 & 66.00  & \textbf{30} & 2149  & 8.52      \\
3 & 3 & 3 & \textbf{42}  & 601  & 0.16 & 43           & 196075 & 99.00  & 43          & 20559 & 673.33    \\
4 & 4 & 4 & \textbf{56}  & 854  & 0.24 & 65           & 251960 & 132.00 & 57          & 49355 & 4133.63   \\
5 & 5 & 5 & \textbf{70}  & 1094 & 0.35 & 74           & 336937 & 165.00 & -           &   -   & -           \\
6 & 6 & 6 & \textbf{84}  & 1320 & 0.46 & 92           & 402579 & 198.00 & -           &   -   & -           \\
7 & 7 & 7 & \textbf{98}  & 1559 & 0.57 & 103          & 437961 & 231.00 & -           &   -   & -           \\
8 & 8 & 8 & \textbf{112} & 1807 & 0.74 & 116          & 503814 & 264.00 & -           &   -   & -           \\ \hline
\end{tabular}
}
\label{table_5}
\end{table*}

\begin{table*}[!htbp]
\centering
\renewcommand{\arraystretch}{1.2}
\caption{Scheduling results of the benchmark systems in \cite{Han2015}.}
\scalebox{1.2}{
\begin{tabular}{ccc|rr|rr|rr}
\hline
Instance & \multicolumn{1}{c}{Jobs} & \multicolumn{1}{c|}{Resources}
& \multicolumn{1}{c}{GFBS}       & \multicolumn{1}{c|}{CPU[s]}
& \multicolumn{1}{c}{IHFBS\cite{Meja2017}}      & \multicolumn{1}{c|}{CPU[s]}
& \multicolumn{1}{c}{A$^{\ast}$\cite{Huang2022}} & \multicolumn{1}{c}{CPU[s]} \\ \hline
FMS01 & 10  & 4  & \textbf{284}  & 0.05 & 293           & 40.00   & 293 & 7.61   \\
FMS02 & 20  & 4  & \textbf{549}  & 0.22  & 557           & 80.00   & 557 & 299.02 \\
FMS03 & 40  & 4  & \textbf{1074} & 1.03 & 1087          & 160.00  & - & - \\
FMS04 & 60  & 4  & \textbf{1613} & 2.66 & 1617          & 240.00  & - & - \\
FMS05 & 100 & 4  & \textbf{2677} & 4.22 & \textbf{2677} & 800.00  & - & - \\
FMS06 & 10  & 8  & \textbf{149}  & 0.12  & 150           & 80.00   & - & - \\
FMS07 & 20  & 8  & \textbf{269}  & 0.45 & 273           & 160.00  & - & - \\
FMS08 & 40  & 8  & \textbf{530}  & 1.66 & 531           & 320.00  & - & - \\
FMS09 & 60  & 8  & 796           & 2.15 & \textbf{795}  & 480.00  & - & - \\
FMS10 & 100 & 8  & 1329          & 6.09 & \textbf{1325} & 800.00  & - & - \\
FMS11 & 10  & 12 & \textbf{106}  & 0.18 & \textbf{106}  & 120.00  & - & - \\
FMS12 & 20  & 12 & \textbf{185}  & 0.27 & \textbf{185}  & 240.00  & - & - \\
FMS13 & 40  & 12 & \textbf{366}  & 1.68 & 368           & 480.00  & - & - \\
FMS14 & 60  & 12 & \textbf{531}  & 3.38 & 532           & 720.00  & - & - \\
FMS15 & 100 & 12 & \textbf{893}  & 7.35 & 897           & 1200.00 & - & - \\
FMS16 & 10  & 16 & \textbf{99}   & 0.26 & \textbf{99}   & 160.00  & - & - \\
FMS17 & 20  & 16 & \textbf{146}  & 0.56 & 149           & 320.00  & - & - \\
FMS18 & 40  & 16 & \textbf{266}  & 0.89 & 274           & 640.00  & - & - \\
FMS19 & 60  & 16 & \textbf{398}  & 2.40 & 401           & 960.00  & - & - \\
FMS20 & 100 & 16 & \textbf{663}  & 12.83 & 664           & 1600.00 & - & - \\ \hline
\end{tabular}
}
\label{table_6}
\end{table*}

\subsection{Performance Comparison with Other Methods}
The first set of tested benchmark instances is taken from \cite{Huang2022}. Each instance includes three types of jobs, three robots, and four types of machine, with each type of machine having two units. One job exhibits routing flexibility, while the others do not. The batch sizes for each job range from 1 to 4.
Four  additional larger instances with the same net structure as the other instances are designed, where job batch size are increased from 4 to 8 without any change in the resource capacity.
We compare our results with the A$^{\ast}$ based algorithm presented in \cite{Huang2022} and the iterated hybrid filtered beam search (IHFBS) algorithm presented in \cite{Meja2017}.
The algorithm is executed with various values for $\beta_g$ in the range [20:100] and $\beta_l$ in the range [2:10].
The A$^{\ast}$ search algorithm utilizes an admissible heuristic function and the IHFBS is based on repeatedly invoking the search algorithm, where the best objective function value currently searched serves as the upper bound for the subsequent call.

The makespan of the resulting schedule $F_{max}$, the number of expanded states $\mathcal{N}$, and the CPU time which refers to the amount of time the CPU spent on the algorithm execution, are presented in Table \ref{table_5}. 
We use ``-" to  denote that the solution cannot be obtained in  3600 seconds. From all the tested instances, the objective function values obtained by our developed approach are less than or equal to those obtained by  IHFBS method and A$^{\ast}$ method, while  the number of expanded states of GFBS is significantly smaller than those of A$^{\ast}$ and IHFBS.
Moreover, in the last four instances,  A$^{\ast}$ algorithm is unable to find a solution within an acceptable time, whereas both GFBS and IHFBS succeeded in doing so. Unlike the GFBS algorithm that executes a single iteration, IHFBS algorithm relies on multiple calls to the beam search algorithm, using the objective function value from the current call as an upper bound for subsequent calls. Consequently, the number of candidates to explore increases significantly, and the computational time is higher than  our method.


The second set of tested benchmark instances is taken from \cite{Han2015} that are labeled with FMS01-FMS20. Each instance comprises two  types of jobs, with batch sizes varying from 10 to 100 parts, and four resource types that have differing capacities, which range from 1 to 4.  Simulation results  are presented in Table \ref{table_6}. 
In 80\%  of these instances (16 out of 20), GFBS found better solutions than the other two methods, while 
IHFBS provides better solutions in two instances FMS09 and FMS10. 
Due to the high computational cost, A$^{\ast}$ algorithm is only able to process the first two instances in an acceptable time, and its running time consumption showed a rapid increase with the problem size.The GFBS algorithm for the smallest instance FMS01 does not require more than 1s to obtain results, while for the largest instance FMS20 it requires 13s. For the IHFBS algorithm, the running time ranges from 40s on FMS01 to 1600s on FMS20. In conclusion, based on the construction of the BRG that compresses the state space of P-TPNS while preserving system information, the GFBS algorithm explores fewer states of the system while maintaining high accuracy.

\section{Conclusion and Future Work}\label{conclusion}
This paper proposes a new scheduling method for FMSs based on P-TPNS. First, a given FMS is modeled using P-TPNS. Then, based on a compact representation of the state space called basis reachability graph, an efficient heuristic approach called generation filtered  beam search  is proposed to optimize the makespan. 
Experimental results are conducted
on several benchmark systems to show that the developed method is superior to the existing methods in terms of search efficiency and solution quality. In the future, our aim is to investigate the scheduling problem with more complex requirements, such as resource cost and energy constraints. Furthermore, the design of new heuristic functions will be considered to improve the search strategy.



\bibliographystyle{IEEEtran}
\bibliography{IEEEtest}
\end{document}